\documentclass[%
 reprint,
 superscriptaddress,
 amsmath,amssymb,
 aps,
 prl,
floatfix,
]{revtex4-2}

\usepackage{graphicx}
\usepackage{dcolumn}
\usepackage{booktabs}
\usepackage{makecell}
\usepackage{bm}
\usepackage{lipsum}
\usepackage{hyperref}
\usepackage[separate-uncertainty]{siunitx}
\DeclareSIUnit\years{yrs}
\DeclareSIUnit\pe{PE}
\DeclareSIUnit\years{years}
\DeclareSIUnit\photoelectron{PE}
\DeclareSIUnit\inch{"}
\usepackage{printlen}
\usepackage{wrapfig}

\usepackage[rightcaption]{sidecap}
\sidecaptionvpos{figure}{c}

\newif\ifincludesupplement
\includesupplementtrue 

\begin{document}

\title{Search for Extremely-High-Energy Neutrinos and First Constraints 
on the Ultrahigh-Energy Cosmic-Ray Proton Fraction with IceCube}

\affiliation{III. Physikalisches Institut, RWTH Aachen University, D-52056 Aachen, Germany}
\affiliation{Department of Physics, University of Adelaide, Adelaide, 5005, Australia}
\affiliation{Dept. of Physics and Astronomy, University of Alaska Anchorage, 3211 Providence Dr., Anchorage, AK 99508, USA}
\affiliation{Dept. of Physics, University of Texas at Arlington, 502 Yates St., Science Hall Rm 108, Box 19059, Arlington, TX 76019, USA}
\affiliation{School of Physics and Center for Relativistic Astrophysics, Georgia Institute of Technology, Atlanta, GA 30332, USA}
\affiliation{Dept. of Physics, Southern University, Baton Rouge, LA 70813, USA}
\affiliation{Dept. of Physics, University of California, Berkeley, CA 94720, USA}
\affiliation{Lawrence Berkeley National Laboratory, Berkeley, CA 94720, USA}
\affiliation{Institut f{\"u}r Physik, Humboldt-Universit{\"a}t zu Berlin, D-12489 Berlin, Germany}
\affiliation{Fakult{\"a}t f{\"u}r Physik {\&} Astronomie, Ruhr-Universit{\"a}t Bochum, D-44780 Bochum, Germany}
\affiliation{Universit{\'e} Libre de Bruxelles, Science Faculty CP230, B-1050 Brussels, Belgium}
\affiliation{Vrije Universiteit Brussel (VUB), Dienst ELEM, B-1050 Brussels, Belgium}
\affiliation{Dept. of Physics, Simon Fraser University, Burnaby, BC V5A 1S6, Canada}
\affiliation{Department of Physics and Laboratory for Particle Physics and Cosmology, Harvard University, Cambridge, MA 02138, USA}
\affiliation{Dept. of Physics, Massachusetts Institute of Technology, Cambridge, MA 02139, USA}
\affiliation{Dept. of Physics and The International Center for Hadron Astrophysics, Chiba University, Chiba 263-8522, Japan}
\affiliation{Department of Physics, Loyola University Chicago, Chicago, IL 60660, USA}
\affiliation{Dept. of Physics and Astronomy, University of Canterbury, Private Bag 4800, Christchurch, New Zealand}
\affiliation{Dept. of Physics, University of Maryland, College Park, MD 20742, USA}
\affiliation{Dept. of Astronomy, Ohio State University, Columbus, OH 43210, USA}
\affiliation{Dept. of Physics and Center for Cosmology and Astro-Particle Physics, Ohio State University, Columbus, OH 43210, USA}
\affiliation{Niels Bohr Institute, University of Copenhagen, DK-2100 Copenhagen, Denmark}
\affiliation{Dept. of Physics, TU Dortmund University, D-44221 Dortmund, Germany}
\affiliation{Dept. of Physics and Astronomy, Michigan State University, East Lansing, MI 48824, USA}
\affiliation{Dept. of Physics, University of Alberta, Edmonton, Alberta, T6G 2E1, Canada}
\affiliation{Erlangen Centre for Astroparticle Physics, Friedrich-Alexander-Universit{\"a}t Erlangen-N{\"u}rnberg, D-91058 Erlangen, Germany}
\affiliation{Physik-department, Technische Universit{\"a}t M{\"u}nchen, D-85748 Garching, Germany}
\affiliation{D{\'e}partement de physique nucl{\'e}aire et corpusculaire, Universit{\'e} de Gen{\`e}ve, CH-1211 Gen{\`e}ve, Switzerland}
\affiliation{Dept. of Physics and Astronomy, University of Gent, B-9000 Gent, Belgium}
\affiliation{Dept. of Physics and Astronomy, University of California, Irvine, CA 92697, USA}
\affiliation{Karlsruhe Institute of Technology, Institute for Astroparticle Physics, D-76021 Karlsruhe, Germany}
\affiliation{Karlsruhe Institute of Technology, Institute of Experimental Particle Physics, D-76021 Karlsruhe, Germany}
\affiliation{Dept. of Physics, Engineering Physics, and Astronomy, Queen's University, Kingston, ON K7L 3N6, Canada}
\affiliation{Department of Physics {\&} Astronomy, University of Nevada, Las Vegas, NV 89154, USA}
\affiliation{Nevada Center for Astrophysics, University of Nevada, Las Vegas, NV 89154, USA}
\affiliation{Dept. of Physics and Astronomy, University of Kansas, Lawrence, KS 66045, USA}
\affiliation{Centre for Cosmology, Particle Physics and Phenomenology - CP3, Universit{\'e} catholique de Louvain, Louvain-la-Neuve, Belgium}
\affiliation{Department of Physics, Mercer University, Macon, GA 31207-0001, USA}
\affiliation{Dept. of Astronomy, University of Wisconsin{\textemdash}Madison, Madison, WI 53706, USA}
\affiliation{Dept. of Physics and Wisconsin IceCube Particle Astrophysics Center, University of Wisconsin{\textemdash}Madison, Madison, WI 53706, USA}
\affiliation{Institute of Physics, University of Mainz, Staudinger Weg 7, D-55099 Mainz, Germany}
\affiliation{Department of Physics, Marquette University, Milwaukee, WI 53201, USA}
\affiliation{Institut f{\"u}r Kernphysik, Universit{\"a}t M{\"u}nster, D-48149 M{\"u}nster, Germany}
\affiliation{Bartol Research Institute and Dept. of Physics and Astronomy, University of Delaware, Newark, DE 19716, USA}
\affiliation{Dept. of Physics, Yale University, New Haven, CT 06520, USA}
\affiliation{Columbia Astrophysics and Nevis Laboratories, Columbia University, New York, NY 10027, USA}
\affiliation{Dept. of Physics, University of Oxford, Parks Road, Oxford OX1 3PU, United Kingdom}
\affiliation{Dipartimento di Fisica e Astronomia Galileo Galilei, Universit{\`a} Degli Studi di Padova, I-35122 Padova PD, Italy}
\affiliation{Dept. of Physics, Drexel University, 3141 Chestnut Street, Philadelphia, PA 19104, USA}
\affiliation{Physics Department, South Dakota School of Mines and Technology, Rapid City, SD 57701, USA}
\affiliation{Dept. of Physics, University of Wisconsin, River Falls, WI 54022, USA}
\affiliation{Dept. of Physics and Astronomy, University of Rochester, Rochester, NY 14627, USA}
\affiliation{Department of Physics and Astronomy, University of Utah, Salt Lake City, UT 84112, USA}
\affiliation{Dept. of Physics, Chung-Ang University, Seoul 06974, Republic of Korea}
\affiliation{Oskar Klein Centre and Dept. of Physics, Stockholm University, SE-10691 Stockholm, Sweden}
\affiliation{Dept. of Physics and Astronomy, Stony Brook University, Stony Brook, NY 11794-3800, USA}
\affiliation{Dept. of Physics, Sungkyunkwan University, Suwon 16419, Republic of Korea}
\affiliation{Institute of Basic Science, Sungkyunkwan University, Suwon 16419, Republic of Korea}
\affiliation{Institute of Physics, Academia Sinica, Taipei, 11529, Taiwan}
\affiliation{Dept. of Physics and Astronomy, University of Alabama, Tuscaloosa, AL 35487, USA}
\affiliation{Dept. of Astronomy and Astrophysics, Pennsylvania State University, University Park, PA 16802, USA}
\affiliation{Dept. of Physics, Pennsylvania State University, University Park, PA 16802, USA}
\affiliation{Dept. of Physics and Astronomy, Uppsala University, Box 516, SE-75120 Uppsala, Sweden}
\affiliation{Dept. of Physics, University of Wuppertal, D-42119 Wuppertal, Germany}
\affiliation{Deutsches Elektronen-Synchrotron DESY, Platanenallee 6, D-15738 Zeuthen, Germany}

\author{R. Abbasi}
\affiliation{Department of Physics, Loyola University Chicago, Chicago, IL 60660, USA}
\author{M. Ackermann}
\affiliation{Deutsches Elektronen-Synchrotron DESY, Platanenallee 6, D-15738 Zeuthen, Germany}
\author{J. Adams}
\affiliation{Dept. of Physics and Astronomy, University of Canterbury, Private Bag 4800, Christchurch, New Zealand}
\author{S. K. Agarwalla}
\thanks{also at Institute of Physics, Sachivalaya Marg, Sainik School Post, Bhubaneswar 751005, India}
\affiliation{Dept. of Physics and Wisconsin IceCube Particle Astrophysics Center, University of Wisconsin{\textemdash}Madison, Madison, WI 53706, USA}
\author{J. A. Aguilar}
\affiliation{Universit{\'e} Libre de Bruxelles, Science Faculty CP230, B-1050 Brussels, Belgium}
\author{M. Ahlers}
\affiliation{Niels Bohr Institute, University of Copenhagen, DK-2100 Copenhagen, Denmark}
\author{J.M. Alameddine}
\affiliation{Dept. of Physics, TU Dortmund University, D-44221 Dortmund, Germany}
\author{N. M. Amin}
\affiliation{Bartol Research Institute and Dept. of Physics and Astronomy, University of Delaware, Newark, DE 19716, USA}
\author{K. Andeen}
\affiliation{Department of Physics, Marquette University, Milwaukee, WI 53201, USA}
\author{C. Arg{\"u}elles}
\affiliation{Department of Physics and Laboratory for Particle Physics and Cosmology, Harvard University, Cambridge, MA 02138, USA}
% ...

\collaboration{IceCube Collaboration}
\email{analysis@icecube.wisc.edu}

\date{May 7th 2025}

\begin{abstract}
We present a search for the diffuse extremely-high-energy neutrino 
flux using \num{12.6} years of IceCube data. 
The nonobservation of neutrinos with energies well above
\SI{10}{\peta\electronvolt} constrains the all-flavor 
neutrino flux at \SI{e18}{\electronvolt} to a
level of $E^2 \Phi_{\nu_e + \nu_\mu + \nu_\tau} \simeq \SI{e-8}{\giga\electronvolt \per\centi\metre\squared \per\second \per\steradian}$, the most stringent limit to date.
Using these data, we constrain the proton fraction of ultrahigh-energy 
cosmic rays (UHECRs) above $\simeq \SI{30}{\exa\electronvolt}$ 
to be $\lesssim$\SI{70}{\percent} (at \num{90}\% CL) 
if the cosmological evolution of the sources is comparable to or stronger 
than the star formation rate. 
This is the first result to disfavor the ``proton-only" hypothesis for UHECR
in this evolution regime using neutrino data.
This result complements direct air-shower measurements by being insensitive
to uncertainties associated with hadronic interaction models.
We also evaluate the tension between IceCube's nonobservation and the $\sim$\SI{200}{\peta\electronvolt} KM3NeT neutrino candidate (KM3-230213A), finding it to be $\sim 2.9 \sigma$ based on a joint-livetime fit between neutrino datasets.

\end{abstract}

\maketitle


\noindent \textit{Introduction}---Extremely-high-energy neutrinos 
(EHE$\nu$, $E_{\nu} \gtrsim \SI{e16}{\electronvolt} = \SI{10}{\peta\electronvolt}$) 
are unique messengers of the distant Universe. 
Unlike photons and ultrahigh-energy cosmic rays 
(UHECRs, $E_{\mathrm{CR}} \geq \SI{e18}{\electronvolt} = \SI{1}{\exa\electronvolt}$), 
neutrinos are chargeless and only weakly interacting,
allowing them to travel undeflected by magnetic fields 
and unattenuated by interactions with background photons. 
Their fluxes are closely linked to the properties of UHECR sources, which
remain unidentified~\cite{10.1093/ptep/ptaa104, Ahlers:2015bma}.
Of particular interest is the chemical composition of UHECRs,
which carries more information about the source environments 
than spectral measurements. Inside source environments, 
UHECRs can interact with ambient photon fields and matter, 
producing \textit{astrophysical neutrinos} carrying 
up to $\sim$\SI{5}{\percent} of the parent cosmic ray energy. 
Additionally, after escaping their sources, 
UHECRs can interact with the cosmic microwave background (CMB) 
and extragalactic background light (EBL), creating a 
flux of \textit{cosmogenic neutrinos}~\cite{berezinsky, Hill:1983xs, 
Engel:2001hd, Hooper:2004jc, Anchordoqui:2007fi, Takami:2007pp, 
ahlers_2009, ahlers_gzk, Kotera:2010yn, yoshida_analytic, 
ahlers_minimal, Aloisio:2015ega, heinze, Romero-Wolf:2017xqe, 
AlvesBatista:2018zui, Heinze:2019jou, moller, vanvliet}. 
Interactions with the CMB are presumed responsible 
for the ``Greisen-Zatsepin-Kuz'min (GZK) cutoff" of extragalactic UHECRs~\cite{greisen, zatsepin_kuzmin}
above $\sim10^{19.6} \,$\si{\electronvolt}.
The production of cosmogenic neutrinos depends on a few key features of UHECR sources: 
their composition, spectrum, and distribution as a function of redshift.
Thus, the measurement or even nonobservation of cosmogenic neutrinos 
can constrain some of these features.

In this Letter, we report a search for EHE$\nu$ using \SI{12.6}{\years} 
of data from the IceCube Neutrino Observatory.
The data were taken between June 2010 and June 2023,
corresponding to 4605 days of livetime.
This is \num{50}\% more exposure than that of the previous IceCube search~\cite{ehe_9yr}.
Additionally, the event selection has been reoptimized,
improving the effective area by $\sim\num{15}\%$ near \SI{1}{\exa\electronvolt}.
The null observation  of cosmogenic neutrinos places significant constraints 
on the cosmological evolution of UHECR sources and, moreover, the composition of UHECRs.
In this work, we investigate the specific hypothesis of a proton-only composition 
of UHECRs with the GZK cutoff generating the
observed suppression of UHECRs at \si{\exa\electronvolt} energies.
The method was proposed in~\cite{ahlers_2009}, 
and in a similar fashion applied to the neutrino measurement by the
Pierre Auger Collaboration~\cite{auger_pfrac}.
We find, at \num{90}\% CL, that the observed fraction 
of UHECRs that are protons at Earth above $\simeq \SI{30}{\exa\electronvolt}$ 
cannot exceed \num{70}\% if the source evolution is comparable to the star formation rate (SFR)---a general tracer of matter density in the Universe. 
These constraints are complementary to, and agree with, 
direct air-shower measurements~\cite{auger_composition, ta_isotropy} by being insensitive to uncertainties associated with hadronic interaction models.

\textit{Data Sample}---IceCube~\cite{icecube_instrumentation} is a 
neutrino detector at the South Pole. 
It consists of \num{5160} digital optical modules (DOMs), 
distributed on \num{86} strings, instrumenting a cubic kilometer of ice 
at depths between \num{1450} and \SI{2450}{\metre}. 
Each DOM hosts a 10-inch photomultiplier tube~\cite{icecube_pmt} 
and readout electronics~\cite{icecube_daq}. 
Charged particles produced in neutrino interactions give rise to Cherenkov 
light when propagating through the ice.
Those Cherenkov photons are detected by DOMs and converted 
into photoelectrons (\si{\pe}). 
On top of the IceCube strings, a surface array called IceTop~\cite{icetop} 
measures cosmic-ray air showers.
EHE$\nu$ events in IceCube are observed as either tracks---light depositions along the trajectory of a long-range $\mu$/$\tau$ produced 
in $\nu_{\mu}$/$\nu_{\tau}$ charged-current interactions---or cascades---approximately spherical light depositions arising from all-flavor 
neutral-current interactions and charged-current interactions of $\nu_e$.

This search aims to select cosmogenic neutrinos while rejecting backgrounds.
Because $\gtrsim100$\,\si{\tera\electronvolt} neutrinos
are absorbed by Earth~\cite{csms},  EHE$\nu$s are expected to be 
downgoing or horizontal at IceCube.
The dominant background is downgoing atmospheric
muon bundles produced in cosmic-ray air showers.
This flux is modeled using 
\textsc{corsika}~\cite{corsika}, with \textsc{sibyll}2.3c~\cite{sibyll2.3c} 
as the hadronic interaction model and the cosmic-ray flux 
prediction from~\cite{Gaisser_h3a}. 
Further backgrounds arise from atmospheric and astrophysical neutrinos.
Atmospheric neutrinos are produced by meson decays during cosmic-ray air showers. 
Their flux is divided into a conventional component~\cite{honda2006} 
originating from pion and kaon decays, 
and a yet-unobserved prompt component~\cite{berss} produced by heavier, short-lived mesons.
All neutrinos---atmospheric, astrophysical, and cosmogenic---are simulated using the \textsc{juliet} code~\cite{juliet}.

The \SI{3}{\kilo\hertz} event trigger rate in IceCube 
is dominated by atmospheric muons, while the cosmogenic neutrino flux 
is already constrained to $\ll 1$ event per year~\cite{ehe_9yr}.
The signal-to-noise ratio is improved by employing an 
event selection based on quality cuts of high-energy events 
in combination with an IceTop veto.
Full details of the event selection are presented in End Matter.
In brief, cosmogenic signal events have extremely high energies
and, therefore, produce large amounts of charge (\si{\pe}). 
As the atmospheric muon background is exclusively downgoing, 
we reject most backgrounds with a zenith-angle-dependent charge threshold.
To do this, the event direction is reconstructed 
with a maximum-likelihood reconstruction using an infinite-length 
track hypothesis~\cite{splinempe}.

Energy loss profiles for single muons show large stochastic variations
as the muons propagate. In contrast, in high-multiplicity muon bundles,
these single-muon fluctuations partly average out.
To leverage this, the energy loss profile of events is
reconstructed along the track direction, 
and stricter charge requirements are imposed upon less stochastic events.
Using stochasticity information improved the effective area
by \num{15}\% between \SI{100}{\peta\electronvolt} 
and \SI{1}{\exa\electronvolt} relative to the previous search.
Lastly, IceTop is used to further reduce the background
from atmospheric events, as described in~\cite{ehe_prl}.
The sample is divided into subsamples of tracks and cascades based
on the reconstructed particle velocity, and their deposited energy
and arrival direction are reconstructed with likelihood-based 
methods~\cite{reco_paper, IceCube:2024csv} 
(cf. End Matter). 

After the event selection, the expected atmospheric background 
is $\num{0.40} \pm 0.03$ events, 
and up to $\sim5$ cosmogenic neutrinos are expected for the 
most optimistic model~\cite{ahlers_gzk}, consisting of \num{73}\% 
tracks and \num{27}\% cascades.
The flux beyond \si{\peta\electronvolt} energies is not well 
constrained, and the expectation strongly depends on the assumed model.
The astrophysical expectation ranges from $\sim9$ events for an unbroken power
law with a hard spectral index ($\gamma = 2.37$)~\cite{diffuse_numu},
down to $\sim 0.5$ events assuming a power law with a cutoff
($\gamma = 2.39$, $E_{\mathrm{cutoff}} = \SI{1.4}{\peta\electronvolt}$~\footnote{These best-fit parameters follow from the analysis scheme in~\cite{globalfit_icrc},
and they will be published in future work.})\nocite{globalfit_icrc}. At the highest energies, $E_{\nu} > \SI{100}{\peta\electronvolt}$, this expectation is reduced to \num{0.9} events and \num{3e-30} events, respectively.

For both astrophysical and cosmogenic neutrinos, a flavor ratio 
of $\nu_{\mathrm{e}} : \nu_\mu : \nu_\tau = 1 : 1 : 1$ at 
Earth is assumed~\cite{nutau_doug}, as well as equal fluxes 
of neutrinos and antineutrinos.

Three events with \si{\peta\electronvolt} energies survive the event selection:
a through-going track~\cite{kloppo}, an uncontained cascade~\cite{hydrangea}, 
and a starting track~\cite{ic190331a}.

\textit{Analysis and Results}---To infer physics parameters, the data are fit using a binned Poisson
likelihood in the space of reconstructed direction and energy, 
following the method in~\cite{ehe_9yr}.
Being in the regime of small statistics, all hypothesis tests and limits are based on ensembles of pseudoexperiments.
Confidence intervals are determined using the likelihood ratio test statistic~\cite{feldman_cousins}.

Systematic uncertainties are treated similarly to~\cite{nutau_doug}: 
They are varied in pseudoexperiments based on estimated priors.
The effect of incorporating systematics is modest, with the differential limit weakened by \num{4}\% at \SI{100}{\peta\electronvolt}, reducing toward \num{1}\% at the highest energies. Details of the likelihood and systematics treatment are available in Supplemental Material~\cite{supplement}.

\nocite{profile_fc,cteq,Yanez:2023lsy,clsim,photonics,lpm_effect,Migdal:1956tc,Gerhardt:2010bj}

\begin{figure}[b]
    \includegraphics[width=\columnwidth]{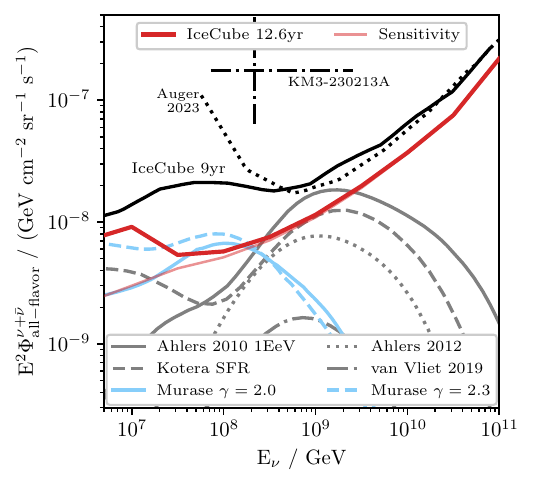}
    \caption[]{
        Differential upper limit (\num{90}\% CL) on the neutrino  
        flux. The differential limit is compared 
        to the IceCube 9 year result~\cite{ehe_9yr}, 
        the limit by Auger~\cite{auger_limit}, the flux inferred from KM3-230213A~\cite{KM3NeT:2025npi},
        cosmogenic neutrino flux 
        models~\cite{ahlers_gzk, Kotera:2010yn, ahlers_minimal, vanvliet} 
        and a UHE astrophysical model~\cite{murase_agn}.
        The model from vV2019~\cite{vanvliet} 
        assumes $\alpha = 2.5$, $E_{\mathrm{max}} = \SI{e20}{\electronvolt}$, $m = 3.4$, 
        and a \num{10}\% proton fraction.
        The Auger limit is rescaled to all-flavor, decade-wide bins for comparison.
    }
    \label{fig:diff_limit}
\end{figure}

\textit{Differential limit and model tests}---The differential upper limit on the neutrino flux above 
\SI{5e6}{\giga\electronvolt} is depicted in Fig.~\ref{fig:diff_limit} 
as a red line. 
The sensitivity, i.e., the limit in case of a null observation, 
and previous limits are also shown.
The limit is weakened with respect to the sensitivity below \SI{100}{\peta\electronvolt}
due to the observed events.
At the highest energies the improvement to the previous limit (IceCube 9yr~\cite{ehe_9yr}) is comparable to the increase in detector livetime as expected for a largely background-free analysis.
At around 1 EeV, additionally, event selection enhancements make a sizable contribution.
Notably, by a few hundred \si{\peta\electronvolt}, the previous limit deviates from the sensitivity due to observed \si{\peta\electronvolt} neutrinos.
In this analysis, although a similar number of \si{\peta\electronvolt} neutrinos were observed, improved energy reconstructions mean they are incompatible with a flux centered at \SI{100}{\peta\electronvolt} and above.

The differential limit is compared to a representative variety of
cosmogenic neutrino models as gray lines.
Qualitatively, larger normalizations, higher maximum accelerating energies, 
and stronger source evolutions generate larger cosmogenic neutrino 
fluxes~\cite{yoshida_analytic, Kotera:2010yn, decerprit}.
In contrast, when the injected cosmic-ray primaries are heavy nuclei, 
photodisintegration becomes the dominant process over photopion 
production and the neutrino flux is suppressed~\cite{Hooper:2004jc, Ave_2005, ahlers_minimal}.
All model predictions shown in Fig.~\ref{fig:diff_limit} 
(except Van Vliet \textit{et al.} 2019~\cite{vanvliet}, abbreviated ``vV2019" hereafter) assume a pure-proton composition with moderate source redshift evolutions comparable to the SFR.
The maximum acceleration energy varies between \SI{e11}{\giga\electronvolt} 
and \SI{e12}{\giga\electronvolt}.

For each aforementioned model, 
we performed a likelihood ratio test as described in Supplemental Material~\cite{supplement}; the results are in Table~\ref{tab:model_tests}.
Although three events were observed, 
the best-fit normalization for a cosmogenic flux component is zero for all tested models.
This indicates the data can be sufficiently explained by astrophysical neutrinos.
All tested cosmogenic models assuming a pure proton composition of UHECRs 
are rejected at \num{95}\% CL.
This indicates that regardless of the differences between those models, 
if the SFR is driving the source evolution of UHECRs, 
a proton-only composition can be excluded.

\newcolumntype{C}[1]{>{\centering\arraybackslash}m{#1}}

\begin{table}[ht]
\caption{
        A selection of cosmogenic neutrino models, 
        the model rejection factor (MRF~\cite{mrf_paper}) at 90\% CL, and associated $p$ value. 
        The analysis strongly $(p<0.05)$ constrains several 
        previously allowed models of the cosmogenic neutrino flux. 
        Cosmogenic models assuming a proton-only composition are marked with a star.
     }
\begin{tabular}{p{4.15cm}C{2.6cm}C{1.4cm}}
\hline \hline
\multicolumn{1}{l}{Model}    & MRF (90\% CL) & $p$ value      \\ \hline
Ahlers 2010$^*$~\cite{ahlers_gzk} (\SI{1}{\exa\electronvolt})               & 0.28 & 0.003 \\
Ahlers 2012$^*$~\cite{ahlers_minimal}                                       & 0.65  & 0.043 \\
Kotera SFR$^*$~\cite{Kotera:2010yn}                                        & 0.49  & 0.027 \\
\makecell[l]{van Vliet~\cite{vanvliet} \\ ($f_p=0.1, m=3.4, \alpha=2.5$)}         & 2.72  & 0.268 \\
\makecell[l]{Murase AGN~\cite{murase_agn} \\ ($\gamma = 2.0, \xi_{\mathrm{CR}} = 3$)}       & 0.47  & 0.057 \\
\makecell[l]{Murase AGN~\cite{murase_agn} \\ ($\gamma = 2.3, \xi_{\mathrm{CR}} = 100$)}     & 0.62  & 0.019 \\ \hline \hline
\end{tabular}
\label{tab:model_tests}
\end{table}

\textit{Proton fraction constraints}---Given the measured UHECR flux, the nonobservation of neutrinos
imposes constraints on the sources. 
This approach is complementary to many existing models, 
which focus on accurately describing the cosmic-ray energy spectrum 
and composition and, thus, also obtain an estimation of the 
accompanying cosmogenic neutrino flux~\cite{heinze, moller, muzio, ehlert}.

The \textsc{crp}{\small{ropa}} package~\cite{crpropa3.2} is used to
model cosmogenic fluxes (following vV2019~\cite{vanvliet}).
In the simulation, protons and secondary neutrinos are propagated
to Earth including energy losses from photopion production
and pair production on the CMB and EBL~\cite{franceschini2008}, 
neutron decay and cosmological adiabatic losses.
Identical sources are distributed homogeneously and isotropically 
with a power-law injection spectrum
$\Phi(E) \propto E^{-\alpha} \exp(-E/E_{\mathrm{max}})$ with spectral index $\alpha \in [\num{1.0}, \num{3.0}]$ 
and exponential cutoff at $E_{\mathrm{max}} \in [\SI{4e10}{\giga\electronvolt}, \SI{e14}{\giga\electronvolt}]$.

Two different models for cosmological source evolution are tested:
\begin{equation}
    \mathrm{SE}_1(z) =
    \begin{cases}
        (1 + z)^m, &z \leq z' \\
        (1 + z')^m, &z > z'
    \end{cases}    
\end{equation}
with $z' = 1.5$ up to $z_{\mathrm{max}} = 4$~\cite{vanvliet}, 
and a more conservative model of $\mathrm{SE}_2(z) = (1+z)^m$ 
with $z_{\mathrm{max}} = 2$, where $m$ denotes the source evolution parameter.
The simulation is normalized to the 
all-particle cosmic-ray flux measured by Auger at 
$10^{10.55} \,$\si{\giga\electronvolt}.
We normalize to the highest-energy data point below the observed
GZK suppression, such that the cosmic-ray flux at the suppression energy is saturated. 
This defines the flux corresponding to a proton fraction
at Earth ($f_p$) of \num{100}\% above energies of $\simeq \SI{30}{\exa\electronvolt}$.
The reference energy impacts the resulting neutrino fluxes.
A systematic shift to the reference energy
on the order of the systematic energy scale of Auger of
$\pm \num{14}\%$~\cite{auger_composition} results in 
a \num{5}\% shift of the overall neutrino flux.

\begin{figure}[ht!]
    \includegraphics[width=\columnwidth]{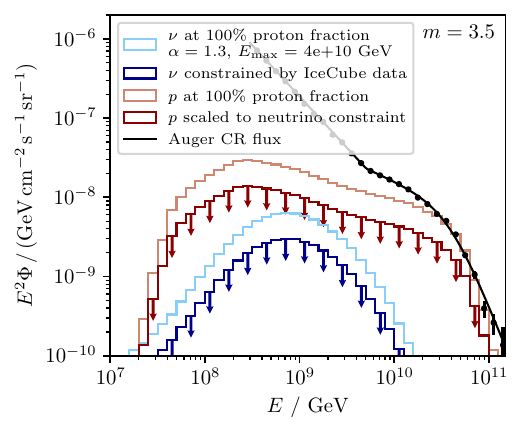}
    \caption[]{
        Illustration of the construction of proton fraction constraints. 
        The red and blue histograms are \num{90}\% CL upper limits 
        on the flux of UHECR protons and cosmogenic neutrinos (per flavor), 
        derived from the nonobservation of EHE$\nu$, assuming
        $\mathrm{SE}_1(z)$ with $m = 3.5$. Also plotted in black is the 
        cosmic-ray flux measured by Auger~\cite{auger_data}. 
        The light-colored histograms represent the case where the proton 
        flux (light red) is allowed to saturate the Auger measurement at 
        $10^{10.55} \, \si{\giga\electronvolt}$.
        The corresponding neutrino flux (light blue) is in tension 
        with the nonobservation in IceCube data and is, therefore, excluded.
    }
    \label{fig:crpropa_sim}
\end{figure}

Figure~\ref{fig:crpropa_sim} shows the construction of the $f_p$ constraints.
The light-colored histograms show the simulated proton flux saturating
the Auger measurement and the secondary neutrino flux.
The source parameters $\alpha$ and $E_{\mathrm{max}}$ are chosen
to minimize the integral neutrino energy flux to obtain a
conservative prediction for a given value of $m$.
The relatively wide range for $\alpha$ is motivated by both experimental and theoretical work, e.g., the Auger Collaboration~\cite{auger_cena} showing that $\alpha$ between 1 and 2 are allowed. Allowing this wider range gives a slightly more conservative result than bounding $\alpha$ at 2.
The range for $E_{\mathrm{max}}$ is bracketed by the ``GZK-cutoff" energy at the low end, and by a value much higher than the observed cosmic rays on the other. In practice, when marginalizing, the minimum value for $E_{\mathrm{max}}$ is always chosen.
The flux shown in the figure is in tension with IceCube data, 
and, thus, $f_p$ can be constrained based on the determined upper limit.
As suggested in vV2019~\cite{vanvliet}, $f_p$ can be determined by comparing the predicted neutrino flux  with the experimental limit at \SI{1}{\exa\electronvolt}.
However, we instead perform a model test, which improves the sensitivity.

\begin{figure}[ht!]
    \includegraphics[width=\columnwidth]{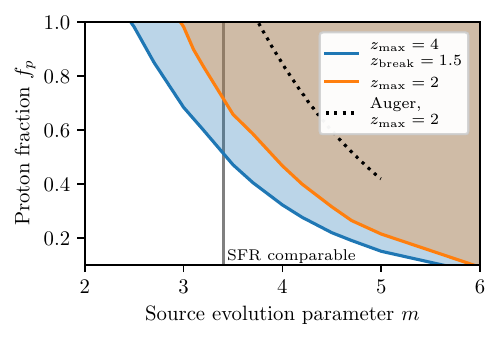}
    \caption[]{
        Constraints on the proton fraction ($f_p$) of UHECRs as a 
        function of source evolution parameter $m$ at \num{90}\% CL 
        based on the nonobservation of UHE neutrinos in this study. 
        The excluded region is shown for the two source evolution models 
        $\mathrm{SE}_1(z)$ (blue) and $\mathrm{SE}_2(z)$ (orange) and
        compared to constraints from Auger~\cite{auger_pfrac}.
    }
    \label{fig:proton_frac}
\end{figure}

This procedure is repeated for different values of the
source evolution parameter $m$, and the resulting constraints are
shown in Fig.~\ref{fig:proton_frac} for the source evolution
models $\mathrm{SE}_1(z)$ and $\mathrm{SE}_2(z)$.
The value of $m$ for UHECR sources is unknown, but here we focus on the range in which $f_p$ can be constrained by this analysis.
For instance, given the source evolution is comparable to the SFR or stronger, 
$f_p$ is constrained to be below about \num{70}\%.
Alternatively, due to the degeneracy between $f_p$ 
and $m$, the results can be interpreted
as an upper bound on the source evolution of $m \lesssim 3$
for proton-dominated UHECRs, strengthening the claim
of the previous analysis~\cite{ehe_prl}.

The predicted neutrino fluxes are dominated by distant cosmic-ray sources,
from which high-energy cosmic rays are not expected to survive.
The model presented here assumes that cosmic-ray sources
are distributed homogeneously within the Universe.
This is true at large distances, but due to Earth's position within 
the Local Supercluster and Local Sheet, 
the local density of sources is enhanced.
This leads to a relative reduction of distant sources
and, thus, of the expected neutrino flux.
Including a model of the local overdensity of sources based on the star
formation rate of local galaxies~\cite{auger_composition, local_density_catalogue} 
weakens the neutrino fluxes, and the corresponding proton fraction 
constraints, by about \num{3}\% ($m = 6.0$) to \num{4}\% ($m = 2.0$).
Additionally, a recent study by Auger shows that up to \num{5}\% 
of UHECRs above \SI{40}{\exa\electronvolt} can be
associated with Centaurus A as the dominant local source~\cite{auger_cena}; 
in this case, the $f_p$ constraint becomes weaker by the same fraction.

For context, constraints already exist from direct air-shower measurements. The Auger $X_{\mathrm{max}}$ data~\cite{auger_composition} and the TA event isotropy~\cite{ta_isotropy} both favor small proton fractions.
However, the data allow~\cite{auger_pfrac, PierreAuger:2023xfc}, and some analyses find~\cite{ehlert,muzio,Muzio:2025gbr}, a proton fraction as high as $\sim \num{10}\%$.
Air-shower-based composition measurements, while powerful, are dependent on hadronic interaction models and the associated uncertainties.
As such, independent measurements that do not rely on air-shower observables directly are highly complementary and necessary.
In particular, our result in this study does not rely on observables from air showers and is, therefore, insensitive to the uncertainties associated with hadronic interaction models.
Although the estimation of the atmospheric muon or neutrino background has a dependence on hadronic interaction models,  the influence on the derived constraints is negligible. 

Auger has also used their neutrino data to constrain $f_p$~\cite{auger_pfrac}, similar to Fig.~\ref{fig:proton_frac}.
Here, we substantially improve the constraints.
In particular, a direct comparison with the $\mathrm{SE}_2$ model with $z_{\mathrm{max}}=2$ can be made, where the resulting exclusion contour---the orange shaded region in Fig.~\ref{fig:proton_frac}---is shifted to smaller values of $m$ relative to the Auger result (the black dotted line) by $\sim 1$.
For example, at $m\sim3.4$, which is comparable to the SFR, we find $f_p \lesssim 70\%$ at 90\%CL, where the Auger result is fully compatible with unity.
We note that the IceCube result achieves this improvement despite making very conservative modeling choices, e.g., marginalizing over $\alpha$ and $E_{\mathrm{max}}$.

\textit{AGN model constraints}---In addition, we tested the active-galactic-nuclei (AGN) 
model from~\cite{murase_agn}, instead of a cosmogenic flux model.
The astrophysical neutrino flux described in this model 
cannot be explained by the observed sub-PeV astrophysical neutrinos 
(cf. Fig.~\ref{fig:diff_limit}).
These UHE astrophysical neutrinos are indistinguishable from 
cosmogenic neutrinos event by event.
The neutrino emission is based on observed photon fluxes, 
using phenomenological parameters like the cosmic-ray loading 
factor $\xi_{\mathrm{CR}}$.
The modeled flux scales linearly with $\xi_{\mathrm{CR}}$, 
so the limit (cf. Table~\ref{tab:model_tests}) can be interpreted 
as an upper limit of $\xi_{\mathrm{CR}} \leq 1.4$ 
and $\xi_{\mathrm{CR}} \leq 62$ for assumed CR spectral indexes 
of $\gamma = \num{2.0}$ and \num{2.3}, respectively.
That the resulting MRFs are $< 1$ indicates that inner jets of 
AGN are unlikely to be a dominant source for UHECRs in this model scenario. 

\textit{KM3-230213A} --- Recently, KM3NeT published a $\sim \SI{220}{\peta\electronvolt}$ neutrino candidate~\cite{KM3NeT:2025npi}. The inferred diffuse flux, also shown in Fig.~\ref{fig:diff_limit}, assumes an $E^{-2}$ ranging from \SI{72}{\peta\electronvolt} to \SI{2.6}{\exa\electronvolt} and significantly exceeds the limits presented in this work. With the exposure of this analysis, this flux leads to an expectation of $\sim 70$ events, inconsistent with our nonobservation at $>10\sigma$; a transient source hypothesis could reduce this tension~\cite{Li:2025tqf,KM3NeT:2025npi,KM3NeT:2025bxl,fang2025}. Considering a joint fit between IceCube, Auger, and KM3NeT, the tension in the diffuse hypothesis is significantly reduced~\cite{KM3NeT:2025ccp}. After repeating the joint fit with the IceCube exposure presented here, the probability of the joint fit resulting in one observed event in KM3NeT (with $\mu_{\mathrm{KM3}} = 0.01$ expected events) and no events in both Auger ($\mu_{\mathrm{A}} = 0.3$) and IceCube ($\mu_{\mathrm{IC}} = 0.68$) is $\sim$\num{0.35}\%. The corresponding goodness-of-fit $p$ value determined by the saturated Poisson likelihood test~\cite{baker_cousins} is \num{0.4}\%~(\num{2.9}$\sigma$).

The impact on $f_p$ constraints depends on the neutrino's origin. If produced in a neutrino source environment, the constraints would be unaffected. If cosmogenic~\cite{KM3NeT:2025vut}, a combined analysis will weaken the inferred limits. 

\textit{Summary}---The nonobservation of neutrinos with energies well above \SI{10}{\peta\electronvolt} in \SI{12.6}{\years} of IceCube data 
places the most stringent limit on cosmogenic neutrino fluxes to date, 
reaching a neutrino flux of
 $E^2 \Phi_{\nu_e + \nu_\mu + \nu_\tau} \simeq \SI{e-8}{\giga\electronvolt \per\centi\metre\squared \per\second \per\steradian}$.
Additionally, we provide the strongest constraints on the composition 
of UHECRs obtained by neutrino astronomy, disfavoring proton-only UHECRs 
if their sources are evolving with the SFR or stronger.

\begin{acknowledgments}
\textit{Acknowledgments}---The IceCube Collaboration acknowledges the significant contributions to this manuscript by Brian A. Clark and Maximilian Meier.
The authors gratefully acknowledge the support from the following agencies and institutions:
USA {\textendash} U.S. National Science Foundation-Office of Polar Programs,
U.S. National Science Foundation-Physics Division,
U.S. National Science Foundation-EPSCoR,
U.S. National Science Foundation-Office of Advanced Cyberinfrastructure,
Wisconsin Alumni Research Foundation,
Center for High Throughput Computing (CHTC) at the University of Wisconsin{\textendash}Madison,
Open Science Grid (OSG),
Partnership to Advance Throughput Computing (PATh),
Advanced Cyberinfrastructure Coordination Ecosystem: Services {\&} Support (ACCESS),
Frontera and Ranch computing project at the Texas Advanced Computing Center,
U.S. Department of Energy-National Energy Research Scientific Computing Center,
Particle astrophysics research computing center at the University of Maryland,
Institute for Cyber-Enabled Research at Michigan State University,
Astroparticle physics computational facility at Marquette University,
NVIDIA Corporation,
and Google Cloud Platform;
Belgium {\textendash} Funds for Scientific Research (FRS-FNRS and FWO),
FWO Odysseus and Big Science programmes,
and Belgian Federal Science Policy Office (Belspo);
Germany {\textendash} Bundesministerium f{\"u}r Bildung und Forschung (BMBF),
Deutsche Forschungsgemeinschaft (DFG),
Helmholtz Alliance for Astroparticle Physics (HAP),
Initiative and Networking Fund of the Helmholtz Association,
Deutsches Elektronen Synchrotron (DESY),
and High Performance Computing cluster of the RWTH Aachen;
Sweden {\textendash} Swedish Research Council,
Swedish Polar Research Secretariat,
Swedish National Infrastructure for Computing (SNIC),
and Knut and Alice Wallenberg Foundation;
European Union {\textendash} EGI Advanced Computing for research;
Australia {\textendash} Australian Research Council;
Canada {\textendash} Natural Sciences and Engineering Research Council of Canada,
Calcul Qu{\'e}bec, Compute Ontario, Canada Foundation for Innovation, WestGrid, and Digital Research Alliance of Canada;
Denmark {\textendash} Villum Fonden, Carlsberg Foundation, and European Commission;
New Zealand {\textendash} Marsden Fund;
Japan {\textendash} Japan Society for Promotion of Science (JSPS)
and Institute for Global Prominent Research (IGPR) of Chiba University;
Korea {\textendash} National Research Foundation of Korea (NRF);
Switzerland {\textendash} Swiss National Science Foundation (SNSF).

\textit{Data availability}---The data that support the findings of
this Letter are not publicly available upon publication
because it is not technically feasible and/or the cost of
preparing, depositing, and hosting the data would be
prohibitive within the terms of this research project. The
data are available from the authors upon reasonable request.
\end{acknowledgments}

\bibliography{refs}

\appendix
\onecolumngrid
\begin{center}
\rule{0.5\textwidth}{0.5pt}
\end{center}
\section*{End Matter}
\twocolumngrid

\setcounter{equation}{0}
\renewcommand{\theequation}{EM-\arabic{equation}}
\textit{Event Selection}---The event selection approach is based on a previous IceCube study~\cite{ehe_prl}, where signal candidates are found by applying four consecutive steps that are designed to remove atmospheric and astrophysical backgrounds.
Table~\ref{tab:cut_table} provides the expected number of background events passing each cut stage, along with the expectation for a cosmogenic neutrino flux~\cite{ahlers_gzk}.

\begin{table}[hbt!]
\caption{For the four analysis cuts, the table describes the number of atmospheric muons, atmospheric neutrinos, and astrophysical neutrinos~\cite{globalfit_icrc} passing the cuts. The final column provides a range of cosmogenic neutrino flux predictions between vV2019~\cite{vanvliet} and~\cite{ahlers_gzk}.}
\begin{tabular}{p{3.2cm}p{1.25cm}p{1cm}p{1.2cm}p{1.25cm}}
\hline \hline
Cut stage             & Atm $\mu$          & Atm $\nu$ & Astro $\nu$ & Cosmo $\nu$ \\ \hline
(1) Charge and hit cut & $5.5 \times 10^4$  & 4.8       & 37          & 2.6--11.5  \\ 
(2) Track quality cut & $8.2 \times 10^3$  & 0.4       & 1.3         & 1.4--8.5   \\ 
(3) Muon bundle cut   & 0.6                & 0.2       & 0.5         & 0.8--5.6   \\ 
(4) IceTop veto       & 0.2                & 0.2       & 0.5         & 0.8--5.4   \\ \hline \hline
\end{tabular}
\label{tab:cut_table}
\end{table}

In the first step of the event selection, only events with a total recorded charge of $Q_{\mathrm{tot}} \geq \SI{27500}{\photoelectron}$ and a number of hit DOMs of $n_{\mathrm{DOMs}} \geq 100$ are kept.
This cut already rejects a majority of atmospheric neutrinos, reducing the expected background to $<10$ events.

\begin{SCfigure*}[0.35]
    \begin{wide}
    \includegraphics[width=0.70\textwidth]{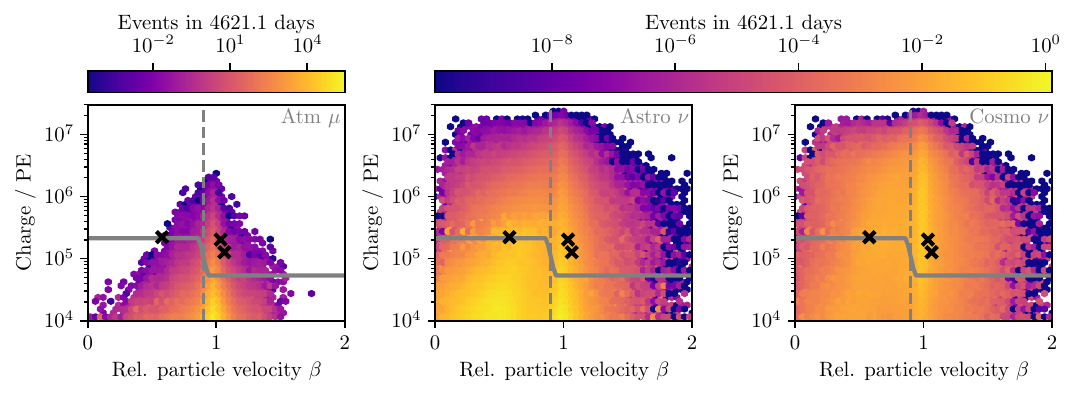}
    \caption{2D histograms of the second stage of the event selection. Distribution of charge vs reconstructed relative particle velocity $\beta$ for for atmospheric muons (left), astrophysical neutrinos (center~\cite{diffuse_numu}), and cosmogenic neutrinos (right~\cite{ahlers_gzk}). The cut applied is shown as a gray dashed line. The three candidate events passing all cuts are shown as black crosses.}
    \label{fig:L3_cut}
    \end{wide}
\end{SCfigure*}

\begin{SCfigure*}[0.45][ht!]
    \begin{wide}
    \includegraphics[width=.65\textwidth]{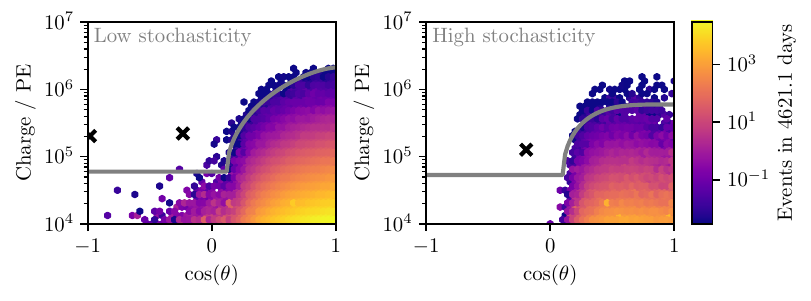}
    \caption{A 2D histogram showing the third stage of the event selection. Both plots show the distribution of charge as a function of reconstructed zenith for atmospheric muons, with the left panel showing low-stochasticity events and the right panel showing high-stochasticity events. The three candidate events passing all cuts are shown as black crosses.
    }
    \label{fig:L4_cut}
    \end{wide}
\end{SCfigure*}

The second step of the event selection is shown as the gray line in Fig.~\ref{fig:L3_cut}. 
The cut is a two-dimensional cut in the plane of reconstructed relative particle velocity $\beta = | \vec{v} | / c$ and the total recorded charge $Q_{\mathrm{tot}}$.
Mathematically,

\begin{equation}
    \hspace{-1em}
    \log_{10}\left(\frac{Q_{\mathrm{tot}}} {\si{\photoelectron}}\right)>\begin{cases}
        5.33 & \beta \leq 0.867 \\
        5.33-30(\beta-0.867) & 0.867 < \beta \leq 0.934 \\
        4.73 & \beta > 0.934.
    \end{cases}
\end{equation}

The cut has multiple purposes. It rejects atmospheric neutrinos, and also rejects misreconstructed atmospheric muon events and neutrino events. 
The speed is reconstructed with the ``LineFit" algorithm~\cite{linefit}, which assumes a light source traveling with a velocity $\vec{v}$ along an infinite-length track. 
For a well-reconstructed track the speed will be distributed around the speed of light ($| \vec{v} | \simeq c \simeq \SI{0.3}{\meter\per\nano\second}$). 
(Apparently ``superluminal" tracks are also possible due to uncertainty of the reconstruction.) At this stage, \num{65}\% of signal events are tracks well reconstructed with $\beta$ within \num{10}\% of $c$, and the majority of outliers are cascades with $\beta < \num{0.9}$.
As a consequence, the speed can also be used to separate the final event sample into subsets of cascades and tracks, which is done at $| \vec{v} | = \SI{0.27}{\meter\per\nano\second}$, shown as a vertical dashed line in Fig.~\ref{fig:L3_cut}.
The design of the event selection is mainly motivated by the distribution of the dominant tracklike events but is applied in the same fashion to all events, including cascades.
After the track quality cut, the atmospheric neutrino expectation is $<1$.
After this cut stage, the sample is dominated by downgoing atmospheric muon bundles.
Therefore, at this stage in the analysis, we use a one-dimensional fit in observed charge between \SIrange[range-phrase=--,range-units=single]{5e4}{e6}{\photoelectron} to determine the overall normalization of the atmospheric muon flux.

\begin{figure}[t!]
    \includegraphics[width=.48\textwidth]{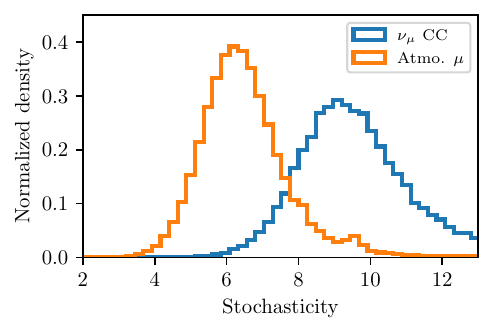}
    \caption[]{Distribution of stochasticity for atmospheric muons (muon bundles) and $\nu_{\mu} \, \mathrm{CC}$ (single muons) events.}
    \label{fig:stochasticity}
\end{figure}

The third step of the event selection is designed to remove the main background of downgoing muon bundles.
The cut is made in the 2D plane of reconstructed particle zenith $\cos(\theta)$ and total recorded charge $Q_{\mathrm{tot}}$ and is visible in Fig.~\ref{fig:L4_cut} for the atmospheric muon background. 
In this plane, the differences between signal (cosmogenic neutrinos) and dominant background (atmospheric muons) appear in both the zenith distribution and the energy loss profile of single muons or taus compared to muon bundles with large muon multiplicities.
As the energy of a muon increases, its energy losses become more stochastic.
In a muon bundle with the same total energy, the energy is distributed among many muons, resulting in a superposition of lower-energy muons losing their energy more continuously, even though their mean $\mathrm{d}E/\mathrm{d}x$ is comparable.
To obtain a measure of the ``stochasticity" of an event, the energy loss profile is reconstructed using a segmented energy loss reconstruction~\cite{reco_paper} over a distance of \SI{40}{\metre}. 
The reconstructed loss profile is then compared to a muon bundle probability density function obtained with \textsc{proposal}~\cite{proposal}. The probability density function is determined by simulating muon bundles for \SI{40}{\metre} repeatedly and recording their total energy loss. Then, we define the reconstructed stochasticity: $\kappa = - \sum_i \log \left[P(\Delta E_i/E)\right] / \mathrm{n.d.f.}$, where the sum runs over all unfolded energy depositions $\Delta E_i$ in the reconstruction, and $\Delta E_i/E$ are the relative energy losses of the event. 
This produces a variable comparable to a reduced log-likelihood, the distribution of which is shown for atmospheric muon bundles and single high-energy muons in Fig.~\ref{fig:stochasticity}.
Events with $\kappa > 8.37$ are regarded as ``highly" stochastic. 

With the goal of removing downgoing muon bundles, the cut imposes a stronger requirement on downgoing events than upgoing events.
The cut is defined by two charge thresholds ($a, b$), a shape parameter $c$, and a transition point from the upgoing to downgoing region $d$:
\begin{equation}
    \log_{10}\left(\frac{Q_{\mathrm{tot}}}{\si{\photoelectron}}\right)>\begin{cases}
        a & \cos(\theta)<d\\
        a + b \sqrt{1 - \Big(\frac{1-\cos(\theta)}{1-d}\Big)^c} & \cos(\theta) \geq d
    \end{cases}
\end{equation}
Parameter values are chosen to maximize the model rejection factor for the cosmogenic neutrino flux prediction in~\cite{ahlers_gzk}.
The final parameter values are $a=4.777$, $b=1.55$, $c=1.5$, and $d=0.12$ for the low-stochasticity events, and $a=4.727$, $b=1.05$, $c=4$, abd $d=0.10$ for high-stochasticity events.
The result is a substantially looser selection for highly stochastic downgoing events, as seen in Fig.~\ref{fig:L4_cut}.
This use of a stochasticity variable is new to this event selection and improves the MRF by more than 10\% relative to the previous event selection.

The fourth and final stage in the event selection uses IceTop to reject atmospheric muons.
IceTop hits correlated with an event in the in-ice detector can be found by extrapolating the reconstructed track to the surface and finding the time $t_{\mathrm{CA}}$, where the track is at its closest approach to IceTop.
Correlated IceTop hits are defined by the collections of hits that satisfy $\SI{-1}{\micro\second} \leq t_{\mathrm{CA}} \leq \SI{1.5}{\micro\second}$.
Events are vetoed if they have two or more correlated hits in IceTop, reducing the remaining atmospheric muon background by $
\sim$\num{60}\% but only reducing the all-sky neutrino rate by $<\num{5}\%$.

\begin{figure}[t!]
    \includegraphics[width=.48\textwidth]{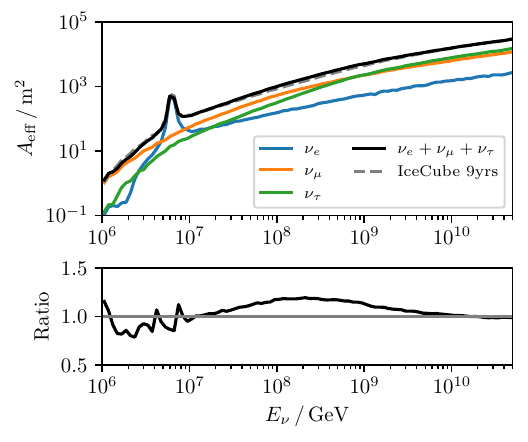}
    \caption[]{The sky-averaged effective area of the analysis as a function of energy. The effective area from the previous iteration of this analysis~\cite{ehe_9yr} is plotted as a dashed line.}
    \label{fig:effective_area}
\end{figure}

The final zenith-averaged neutrino effective area for the event selection (before applying the IceTop veto) is shown in Fig.~\ref{fig:effective_area} and compared to the previous version of the event selection~\cite{ehe_prl}. The effective area describes the neutrino-antineutrino average. The new event selection mostly improves the $\nu_{\mu}$ effective area between \SI{10}{\peta\electronvolt} and \SI{1}{\exa\electronvolt} by about 30\%, while reducing the $\nu_e$ and $\nu_{\tau}$ effective area between \num{1} and \SI{10}{\peta\electronvolt} to reduce the background of astrophysical neutrinos.

\textit{Data Release}---A data release containing the main results of Figs.~\ref{fig:diff_limit} and~\ref{fig:effective_area} is available online~\cite{DVN/JHK49D_2025}.

\onecolumngrid
\begin{center}
\rule{0.5\textwidth}{0.5pt}
\end{center}
\twocolumngrid

\ifincludesupplement

\onecolumngrid
\noindent
\makebox[\textwidth]{%
  \begin{minipage}{0.95\textwidth}
  \centering
  \vspace*{1cm}
  {\large \textbf{Supplemental Material for}}\\[1em]
  {\large ``A search for extremely-high-energy neutrinos and first constraints}\\
  {\large on the ultra-high-energy cosmic-ray proton fraction with IceCube"}\\
  \vspace{0.5cm}
  {\large \textit{The IceCube Collaboration}}
  \vspace{1cm}
  \end{minipage}
}
\\
\twocolumngrid

\setcounter{equation}{0}
\renewcommand{\theequation}{S-\arabic{equation}}

\textbf{Likelihood Construction:}
To infer physics parameters, the data is fit using a binned Poisson
likelihood in the space of reconstructed direction and energy. The likelihood formulation is adopted directly from the previous search~\cite{ehe_9yr}.
The observable binning is likewise adopted from~\cite{ehe_9yr}, 
except here \num{10} directional cascade bins are 
used. The cascade angular binning is guided by the 
resolution of recent reconstruction algorithms~\cite{IceCube:2024csv}.

The expectation in each observable bin $i$ is the sum 
of cosmogenic ($\mu_{\mathrm{GZK}, i}$) and astrophysical ($\mu_{\mathrm{astro}, i}$) 
neutrinos, plus all other atmospheric
backgrounds ($\mu_{\mathrm{bkg}, i}$), and is compared to the
number of observed events $n_i$.
The overall normalizations of the cosmogenic and astrophysical
models---$\lambda_{\mathrm{GZK}}$ and $\lambda_{\mathrm{astro}}$--- 
are allowed to float, with the latter a nuisance parameter.
\begin{multline}
\label{eq:likelihood}
    \mathcal{L}(\lambda_{\mathrm{GZK}}, \lambda_{\mathrm{astro}}) = \\ \prod_i \mathrm{Pois}(n_i | \lambda_{\mathrm{GZK}} \mu_{\mathrm{GZK}, i} + \lambda_{\mathrm{astro}} \mu_{\mathrm{astro}, i} + \mu_{\mathrm{bkg}, i}).
\end{multline}
Being in the regime of small statistics, 
all hypothesis tests are based on ensembles of pseudo-experiments.
Confidence intervals are determined using the likelihood ratio
test statistic (TS)~\cite{feldman_cousins}.

The compatibility of a cosmogenic neutrino model with observed data 
is tested via the likelihood ratio test:
\begin{equation}
    \Lambda = \log \left( \frac{\underset{\lambda_{\mathrm{GZK}}, \lambda_{\mathrm{astro}}}{\sup} \mathcal{L}(\lambda_{\mathrm{GZK}}, \lambda_{\mathrm{astro}})}{\underset{\lambda_{\mathrm{astro}}}{\sup} \mathcal{L}(\lambda_{\mathrm{GZK}} = 1, \lambda_{\mathrm{astro}}) }\right),
    \label{eq:model_test}
\end{equation}
where $\sup$ denotes the supremum.

To determine EHE$\nu$ flux constraints 
in a more model-independent manner, a differential upper limit is constructed.
For this, a sliding $E^{-1}$ neutrino spectrum extending half a decade to both
sides around the central energy is injected in half-decade-wide steps~\cite{ehe_9yr}.
For this flux, a spectral index of $\gamma = 1$ is chosen for comparability with previous results
and results from other experiments.

\textbf{Astrophysical Flux Treatment:} To perform hypothesis tests and calculate upper limits, we must make assumptions about the TeV-PeV astrophysical neutrino flux.
Such assumptions are also needed when constructing pseudo-experiments.

For model-specific hypothesis tests, we assume a power-law spectrum with spectral index $\gamma = 2.37 \pm 0.09$ observed in the energy range \SI{15}{\tera\electronvolt} to \SI{5}{\peta\electronvolt}~\cite{diffuse_numu}, where the indicated range is considered as a systematic uncertainty.
In building pseudo-experiments for these hypothesis tests, the value of $\lambda_{\mathrm{astro}}$ is the best fit obtained from data~\cite{profile_fc}.
This balances model rejection power with the discovery potential for cosmogenic neutrinos, while giving good coverage. 
The effect of assuming a softer spectral index
($\gamma = 2.52$~\cite{globalfit_icrc}) 
ranges from \SI{-30}{\percent} (Murase \cite{murase_agn}) 
to \SI{+6}{\percent} (van Vliet \cite{vanvliet}),
and is negligible at the \SI{1}{\percent}-level for
Ahlers 2010~\cite{ahlers_gzk}.

For construction of the differential neutrino limit and proton fraction constraints, we assume a single power law with exponential cutoff~\cite{globalfit_icrc}. In building pseudo-experiments we assumed $\lambda_{\mathrm{astro}} = 0$. The treatment is different in order to generate conservative upper-limits. This is because we are building upper-limits by construction. That is, even in case of an observation incompatible with background, only upper limits are reported. Of the commonly assumed models for the astrophysical neutrino flux---single-power laws, broken power laws, etc.---we found that assuming a single power-law model with exponential cutoff generated the most conservative results.

The assumptions made in this paper, and discussed above, differ from those made previously in \cite{ehe_9yr}. Where the previous work assumed a relatively hard spectrum of $\gamma=2$, our assumptions here are in better agreement with recent measurements, and produce more conservative results.
A full ``joint-fit" with the TeV-PeV data will be helpful in further improving these constraints and is the topic of future work.

\textbf{Systematic Uncertainties:}
The impact of systematic parameters is estimated
by varying them in pseudo-experiments based on estimated priors.
This procedure modifies $n_i$ in Eq.~\ref{eq:likelihood}, widens the distribution of TS
values, and thus the extracted confidence intervals.
The parameters taken into account are: 
the optical efficiency of the DOMs  (\SI{\pm 10}{\percent})~\cite{icecube_pmt}, 
the neutrino cross section ($^{+3}_{-20}$ \si{\percent})~\cite{csms, cteq},
the average neutrino inelasticity (\SI{\pm 20}{\percent})~\cite{csms, cteq}, 
the atmospheric muon flux ($^{+73}_{-46}$ \si{\percent})
and the conventional (\SI{\pm 30}{\percent})~\cite{Yanez:2023lsy} 
and prompt (\SI{\pm 100}{\percent})~\cite{berss} atmospheric neutrino flux.
The uncertainty on the atmospheric muon flux is dominated
by composition uncertainties of the cosmic-ray flux at Earth
and the range is constructed by re-weighting the simulation
to a primary cosmic-ray flux of protons or
iron nuclei only~\cite{corsika, Gaisser_h3a}.
The baseline normalization of the atmospheric muon component is measured
from a fit to sub-threshold data.
The uncertainty on the cross section has two components -- one from experimental uncertainties in extracting the parton distribution functions (PDFs) which are inputs to the calculation (\SI{\pm 3}{\percent}), and a second theoretical uncertainty in accounting for the effect of heavy quarks on the evolution of the PDFs at high energies according to perturbative QCD.
The baseline cross section value from \cite{csms} shows that the value above $\sim \SI{e4}{\giga\electronvolt}$ can be up to $\sim \SI{20}{\percent}$ \textit{smaller} depending the treatment of the bottom/top quark splitting. This leads to a conservative, asymmetric systematics uncertainty on the cross section of +\SI{3}{\percent} and \SI{-20}{\percent}.

\textbf{Simulation of EHE$\nu$s in IceCube}: Simulation of EHE$\nu$s using the conventional IceCube method of ray-tracing individual photons~\cite{clsim} is computationally infeasible. Instead, pre-calculated photon tables are used~\cite{photonics} to reduce the computational cost. The baseline simulation does not include the Landau-Pomeranchuk-Migdal (LPM) effect~\cite{lpm_effect, Migdal:1956tc, Gerhardt:2010bj}. The scale of the effect was investigated by a special $\nu_e$ simulation including the LPM cascade elongation, and the effect was found to be a sub-percent effect on the total event rate of the EHE$\nu$ sample used in this work. The LPM effect on $\pi^0$-production in hadronic showers was not tested here.
\fi

\end{document}